# A Liquid Density Sensor Based On AlN Piezoelectric Micromachined Ultrasonic Transmitter Insensitive to Liquid Viscosity

Xuecong Fu [1], Liang Lou[2,3]**, and Hao Ren[1]*

[1] School of Information Science and Technology, ShanghaiTech University, Shanghai, 201210, China
[2] School of Microelectronics, Shanghai University, Shanghai 201800, China
[3] Shanghai Industrial μTechnology Research Institute, Shanghai 201899, China

*Corresponding author E-mail: renhao@shanghaitech.edu.cn
**Corresponding author E-mail: liang.lou@sitrigroup.com

**Abstract:**

**To overcome the limitations of conventional liquid density sensors, including bulky size, significant sample requirements, and challenges in automation, MEMS-based approaches have been developed. However, the viscosity–density coupling effect often compromises accuracy in high-viscosity liquids. Although various decoupling strategies have been proposed, they often suffer from complexity and inefficiency. This study presents an AlN-based PMUT liquid density sensor insensitive to viscosity interference. The sensor employs two identical PMUTs, functioning as transmitter and receiver, respectively. An ultrasonic wave generated by the transmitter is reflected by the liquid surface and detected by the receiver. Theoretical calculations demonstrate that when the excitation frequency remains constant, the amplitude of the received electrical signal exhibits a specific relationship with liquid density, while the viscosity-induced signal amplitude variation becomes negligible. Therefore, after calibrating the PMUT liquid density sensor by fitting the relationship between received signal amplitude and density across 0-100% glycerol solutions, the device can measure densities within this range regardless of liquid type and viscosity. Experimental results show that the sensor can accurately measure the density of the propylene glycol solutions using the glycerol-calibrated fitting formula, with the calculated density error rate between measured propylene glycol solutions using the glycerol-calibrated fitting formula and actual densities remains below 0.125%, demonstrating measurement insensitivity to viscosity differences between the two liquids. The proposed method achieves maximum error rates of less than 2.5% in high-viscosity environments (80%-100% glycerol solutions), which is 20% that of other density measurement methods based on resonant frequency. The developed PMUT liquid density sensor exhibits a measurable density range from 0.789 to 1.261 g/cm³.**

**Keywords**: Liquid density, Piezoelectric micromachined ultrasonic transducer (PMUT), Single measurement frequency, Resolution, Resonant frequency.

## 1. Introduction

In many fields, measuring the density of liquids is one of the essential steps. For example, in biomedical applications, the density of body fluids like blood and urine can indicate the health of an individual [1]; in the food industry, detecting liquid density can determine the state of fermentation [2]; in the chemical industry, determining the density of liquids can improve process monitoring and chemical analysis [3]. Although measuring liquid density is crucial, traditional measurement methods, such as the hydrostatic pressure method, buoyancy method, and resonance method, have limitations such as large sensor size, requiring a large amount of liquid samples, manual operation, and unsuitability for automatic measurement [4-6].

To address these technical challenges, miniaturized devices can not only effectively reduce production costs and energy consumption, but also achieve rapid response through micro samples. For this reason, MEMS technology has been rapidly developed and widely adopted in the past half-century, and liquid density sensors based on different sensing principles such as cantilever beam [7], acoustic wave [8], tuning fork [9], and diaphragm [10] have emerged. However, when it comes to the testing of high-viscosity liquid density, the coupling effect of viscosity and density significantly affects the accuracy of the results. For instance, Ghatkesar *et al.* [11] distinguished between liquid density and viscosity by measuring the resonance frequency and quality factor of multiple flexural modes of vibration of a micro-cantilever beam density sensor in glycerol aqueous solution using the Elmer Dreier model. However, it requires precise and independent excitation of the desired two or more vibration modes, and the vibration of different modes may interfere with each other. In addition, measuring multiple parameters, i.e., both quality factor (Q) and resonance frequency in several vibration modes for distinguishing viscosity and density, may introduce more measurement errors, and for higher-order modes with smaller amplitudes, their signals can easily be overwhelmed by the strong fundamental mode signal or noise. Yang *et al.* [12] optimized the structure of a tuning fork liquid density sensor, then used a partial least squares model to fit the experimental data of frequency-density characteristics, and established a density calculation model based on viscosity compensation to obtain more accurate density values by combining frequency-viscosity characteristic experiments. However, each measurement by using this model requires additional determination of the viscosity of the liquid to be tested, which adds complexity to the measurement process. Zhang *et al.* [13] established a fluid dynamics model for quartz tuning fork liquid density sensors to achieve density and viscosity sensing, and obtained density and viscosity by solving equations after measuring multiple resonance frequencies and quality factors. However, in high-viscosity liquids, the quality factor will decrease sharply, causing the resonance peak to be very wide and flat, making it difficult to accurately determine its peak frequency and half-peak width, thus undermining the reliability of f and Q measurements. Herrmann *et al.* [14] reported a micro-acoustic sensor based on a layered quartz/$SiO_2$ system with sagittal corrugations for Love mode, which can realize liquid capture effect through a double delay line device, thus separately measuring liquid density and viscosity. However, although this method can make the response ratio of love mode devices to liquid capture and viscous coupling significantly higher than that of quartz crystal microbalance devices, viscous coupling will still have an impact on density measurement.

Among numerous micro liquid density sensors, the PMUT liquid density sensor has become a research focus due to its unique advantages. Compared with liquid density sensors based on cantilever beam structure, PMUT liquid density sensors have a higher Q, making it easier to distinguish the resonant frequency when working in liquids with higher density, thus improving the sensor's resolution [15, 16]. Compared with CMUT liquid density sensors, PMUT liquid density sensors have advantages like lower bias voltage, good linearity, and low energy loss [17-19]. The resonant frequency of a surface acoustic wave (SAW) sensor is high, which is usually hundreds of megahertz to several gigahertz. The resonant frequency of a PMUT liquid density sensor is lower than that of an SAW sensor, and also has a lower power consumption [20-23]. Compared to the tuning fork structure-based liquid density sensors, PMUT sensors have higher sensitivity [24-27]. For liquid density sensors based on resonators that vibrate in non-surface wave modes, such as the sensor designed by Manzanoque *et al.* [28] using the second-order out-of-plane mode of a piezoelectrically actuated microplate, their anchor structure is easily damaged in harsh environments, leading to sensor failure [28-30].

Many prior studies have attempted to measure the density of liquids by monitoring the resonant frequency and quality factor changes of PMUTs immersed in liquids with different densities. However, the interaction between viscosity and density will affect the resonant frequency and quality factor of PMUT, resulting in deviations in the density measurement of high-viscosity liquids. For example, Ledesma *et al.* [31] demonstrated an AlScN PMUT-CMOS chip with double top electrodes, which showed a linear relationship between the resonant frequency and the liquid density in low-concentration liquids. However, in high-concentration glycerol solutions, the relationship between measured resonant frequency and density was nonlinear, which made this scheme difficult to accurately measure the density of high-viscosity liquids. Roy *et al.* [1] developed a PMUT-fluid-PMUT (PFP) system as a fluid density sensor, which used the linear relationship between the resonant frequency and density in low-viscosity liquids. However, this design required a pair of PMUTs arranged oppositely to each other, making its installation on arbitrary surfaces challenging. They tested the vibration resonance frequency of the device in high-viscosity castor oil (994 kg/m$^3$, 637 cp) and pure water (998 kg/m$^3$, 0.89 cp), and found that the high viscosity of castor oil affected the value of resonance frequency, which illustrated that the liquid density sensor could not be applied to the environment of high-viscosity liquid. Subsequent studies, such as Roy *et al.*'s lead zirconate titanate (PZT) dual-electrode piezoresistance microfluidic sensor integrated microfluidic channel, have solved the problem of PFP systems not being able to be installed on the same surface [32]. However, it still faced the problem of insufficient measurement accuracy of high-viscosity liquid density [33]. Tsao *et al.* [34] used machine learning technology algorithms to characterize liquid properties by analyzing changes in the acoustic signals generated by PMUTs (including parameters such as time of flight, signal amplitude, etc.), thereby obtaining more accurate measurement results and achieving synchronous measurement of fluid viscosity and density. Although this method could simultaneously detect the viscosity and density of liquids, its model structure was relatively complex and was only effective under low-density and low-viscosity conditions.

In this work, we introduced an AlN-based PMUT liquid density sensor only sensitive to liquid density and not sensitive to liquid viscosity variations. Two identical PMUTs placed side by side in the liquid were selected as theas the transmitter and receiver of the liquid density sensor. A single-frequency excitation voltage was applied to drive the transmitter PMUT to generate ultrasonic waves, which were reflected by the liquid surface and subsequently received by the receiver PMUT to produce an electrical signal. Theoretical calculations demonstrate that when the excitation frequency remains constant, the amplitude of the received electrical signal exhibits a specific relationship with liquid density, while the viscosity-induced signal amplitude variation becomes negligible. Therefore, after calibrating the PMUT liquid density sensor by fitting the relationship between received signal amplitude and density across 0-100% glycerol solutions, the device can measure densities within this range regardless of liquid viscosity. In the liquid density range from 1.013 g/cm$^3$ to 1.056 g/cm$^3$, the output voltages of six different density propylene glycol solutions are highly consistent with the calibration curve established by glycerol solution in this density range, and substituting the output voltage value into the fitting equation to calculate the density, the error rate between the calculated density and the real density is smaller than 0.125%. To overcome the constraints of the scanning method for measuring liquid density, we employed a fixed frequency as the excitation signal frequency and derived the liquid's density directly from the output voltage amplitude. Furthermore, the transmitter utilized a 3 V excitation voltage at a fixed frequency, offering a more streamlined excitation circuit compared to the scanning method's circuit. The proposed method achieves maximum error rates of less than 2.5% in high-viscosity environments (80%-100% glycerol solutions), which is 20% that of other density measurement methods based on resonant frequency. Given that the received signal shared the same frequency as the transmitter's excitation signal and boasted a high SNR, the measurement circuit at the receiving end was also straightforward. This dual-PMUT density sensor system achieves a Q of 76.8 in water, surpassing the performance of cantilever beams, tuning fork resonators, and flexural mode resonators. With a cavity radius of up to 500 μm for the PMUT, the output signal exhibits a remarkable SNR of up to 126 dB in pure water. The resolution of the AlN-based PMUT liquid density sensor was evaluated, demonstrating the superior detection resolution of our proposed PMUT liquid density sensor, which could reach up to $2.6 \times 10^{-4}$ g/cm$^3$.

## 2. OPERATION PRINCIPLE

As show in Fig. 1(a), this device is composed of two identical PMUTs, one is the transmitter PMUT and the other is the receiver PMUT, placed side by side in a liquid medium. The AlN-based PMUT consists of an AlN piezoelectric layer sandwiched between two molybdenum metal electrodes, a silicon support layer, and a silicon substrate with a back-etched cavity. The laminated plate of the electrode layer, piezoelectric layer, and support layer forms a vibrating diaphragm, which is released by etching from the back of the substrate. When an alternating voltage is applied between the upper and lower electrodes, the piezoelectric layer deforms, driving the diaphragm to vibrate and emit spherical ultrasonic waves due to the converse piezoelectric effect. Reflected at the interface between liquid and air, ultrasonic waves are received by the receiver PMUT, and an alternating signal due to the piezoelectric effect is generated. As the PMUT is in liquid, the liquid produces an additional mass effect on the vibration of the membrane. At this point, the diaphragm and liquid at the transmitter PMUT can be regarded as a first-order damped harmonic oscillator undergoing forced vibration, which can be analyzed using the mass-spring-damper model [35, 36]. When considering the fluid-solid coupling effect, the spring model of the PMUT in the liquid needs to introduce the effects of fluid inertia and damping on the basis of the traditional mass-spring-damper system [35, 36]. The total external force acting on this model is equal to the inverse piezoelectric force minus the elastic force and resistance. According to Newton's second law:

$$F_{total} = F_{pizeo} - k_{eq}x - b\dot{x} = m_{eff}\ddot{x} \qquad (1)$$

where $F_{total}$, $F_{pizeo}$, $k_{eq}$, $x$, $b$, and $m_{eff}$ are the total force, piezoelectric effect force, spring constant, displacement, damping coefficient, and equivalent mass including additional mass. The resistance $b\dot{x}$ is the resultant force of the mechanical damping force, viscosity damping force, and acoustic radiation damping force [37]. It is the product of the sum of the mechanical damping coefficient $c_{eq}$, viscous damping coefficient $c_{vis}$, and acoustic radiation damping coefficient $c_{rad}$, and the vibration velocity. The elastic force is the product of the elastic coefficient $k_{eq}$ and the displacement $x$. $m_{eff}$ is the total mass of the vibration system. Organizing (1) yields:

$$(m_{eff})\ddot{x} + (c_{eq} + c_{vis} + c_{rad})\dot{x} + k_{eq}x = F_{pizeo} \qquad (2)$$

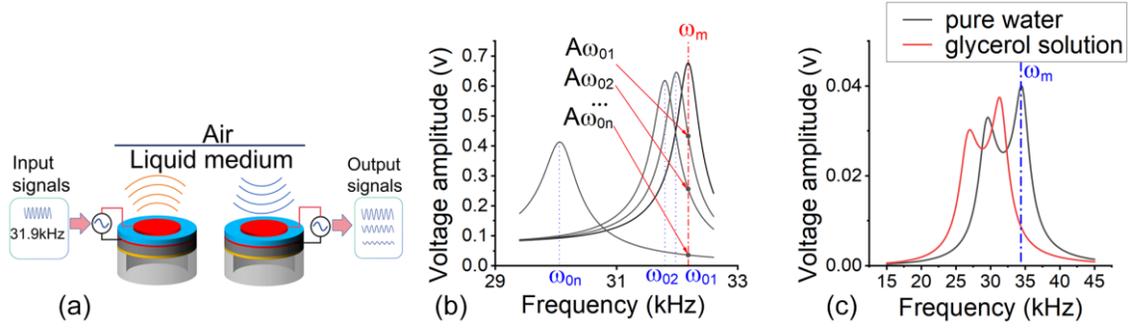

Fig. 1. (a) Schematic diagram of this liquid density sensor. (b) Schema-tic diagram of measuring output voltage in glycerol solutions of different densities using a fixed frequency. (c) The amplitude-frequency charac-teristic curve of the receiver in the case of a mismatch between the resonant frequencies of the transmitter and receiver. The black curve represents the amplitude-frequency characteristic curve of the receiver in pure water; the red curve represents the amplitude-frequency characteristic curve of the receiver in glycerol solution. The horizontal axis of the blue line indicates the selected measurement frequency.

The solution of the amplitude $A$ can be obtained from (2) as:

$$A = \frac{F_0}{\sqrt{[k_{eq}-\omega^2 m_{eff}]^2 + \omega^2(c_{eq}+c_{vis}+c_{rad})^2}} \qquad (3)$$

$$k_{eq} = \omega_{eq}^2 \cdot m_{eq} \qquad (4)$$

where $F_0$ is the amplitude of $F_{pizeo}$, $\omega_{eq}$ is the eigenfrequency, $m_{eq}$ is the PMUT diaphragm mass, and $\omega$ is the vibration angle frequency.

The mechanical damping coefficient, $c_{eq}$, typically characterizes the damping effects arising from internal material friction and energy dissipation at structural joints within the system, that is,

$$c_{eq} = \frac{\gamma k_{eq}}{\omega} \tag{5}$$

where $\gamma$ is the loss factor of silicon ($\gamma=1\times10^{-4}$), and $\omega$ is the vibration angle frequency.

Under resonant vibration conditions, the fluid on the surface of the PMUT membrane vibrates together with the membrane. Therefore, the fluid dynamics scalar parameter is the Stokes number (also known as the dynamic Reynolds number in some literature). The Stokes number represents the ratio between the additional mass force and the viscous damping force. The calculation equation is [38-40]:

$$Stk = \frac{\rho \omega a^2}{\eta} \tag{6}$$

where $a$ is a characteristic length for the particular flow (it is the radius of the PMUT diaphragm), $\eta$ is the dynamic viscosity of the medium, $\rho$ is the density of the medium, and $\omega$ is the vibration angle frequency.

When the Stokes number is much greater than 1 (this device is about 97 in glycerol), it confirms that inertia dominates and the viscous effect can be ignored. At this point, when the circular PMUT film undergoes harmonic vibration in the fluid, the equation for calculating its additional mass is [41]:

$$m_{add} = 0.65\pi\rho a^3 \tag{7}$$

$$m_{eff} = m_{eq} + m_{add} \tag{8}$$

According to Kozlovsky 2009, the quality factor due to viscosity is calculated as [42]

$$Q_{vis} = \frac{0.95}{\xi} \tag{9}$$

Where $\xi$ is calculated as

$$\xi = \sqrt{\frac{\eta}{\rho_f \omega a^2}} \tag{10}$$

where $\eta$ is the dynamic viscosity of the liquid, $a$ is the radius of the diaphragm, $\rho_f$ is the density of the liquid, and $\omega$ is the angular frequency of oscillation.

According to Thomson 1983 [43], the viscous damping coefficient of this PMUT ($c_{vis}$) is calculated by $c_{vis} = {m_{eff}\omega}/{Q}$. Therefore, the viscous damping coefficient in (2) is calculated as:

$$c_{vis} = \frac{m_{eff}\omega}{Q_{vis}} \tag{11}$$

From (7) to (11):

$$c_{vis} = \frac{m_{eff}\omega\sqrt{\eta}}{0.95\sqrt{\rho_f \omega a^2}} = \frac{m_{eff}\sqrt{\eta\omega}}{0.95\sqrt{\rho_f a^2}} \tag{12}$$

For PMUT in liquid, part of its energy is dissipated in the form of sound waves, resulting in damping [37]. This damping is called acoustic radiation damping $c_{rad}$, which is calculated by [44]:

TABLE I
PMUT MATERIAL PROPERTIES.

| Layer | Material | Young's modulus(Gpa) | Density(kg/m³) | Poisson's ratio |
|---|---|---|---|---|
| Top electrode | Mo | 312 | 10200 | 0.31 |
| Piezoelectric | AlN | 410 | 3300 | 0.27 |
| Bottom electrode | Mo | 312 | 10200 | 0.31 |
| Substrate | Si | 166 | 2329 | 0.28 |

TABLE II
WATER-GLYCEROL MIXTURES PROPERTIES

| Property | Glycerol weight percent (%) | | | | | |
|---|---|---|---|---|---|---|
| | 0 | 20% | 40% | 60% | 80% | 100% |
| Density(kg/m³) | 1000 | 1052.2 | 1104.4 | 1156.6 | 1208.8 | 1261 |
| Viscosity(cP) | 0.89 | 1.38 | 2.78 | 7.56 | 36.4 | 648.2 |
| Sound velocity(m/s) | 1482 | 1545 | 1627 | 1719 | 1817 | 1920 |

$$c_{rad} = R_m = 4\pi a^2 c \rho_f \left[\frac{(ka)^2}{1+(ka)^2}\right] \tag{13}$$

where $R_m$ is the real part of the complex form of acoustic radiation impedance $Z_{rad}$ ($Z_{rad}=R_m+jX_m$). According to Kinsler et al. 2000 [44], $R_m$ represents energy dissipation, which equals the acoustic radiation damping coefficient $c_{rad}$. $c$ is wave speed, $k$ is the acoustic radiation wave number, $k=2\pi f/c$, and the acoustic radiation efficiency factor is $ka=2\pi fa/c$. In our experiment, the frequency of the sound wave generated by the sound source vibrating in water is $f$=31.9 kHz, $a$=500×10⁻⁶ m, $c$=1482 m/s, so $ka$=0.0676≪1, the radiation impedance $R_m$ is calculated using a monopole sound source model. The above equation is precisely the calculation equation for the acoustic radiation impedance of a monopole sound source.

Through the above theoretical equations, we can obtain the theoretical value of the damping coefficient of the PMUT when it is vibrated in liquids with different densities. Based on the parameters of the experimental device in this article (Table I and II), we can calculate the $c_{eq}$, $c_{vis}$, and $c_{rad}$ of the device in glycerol solutions of different concentrations using (5), (12), and (13), respectively, as shown in Table III.

TABLE III
DAMPING COEFFICIENT OF LIQUIDS WITH DIFFERENT CONCENTRATIONS

| Glycerol Volume Percent | 0 | 20% | 40% | 60% | 80% | 100% |
|---|---|---|---|---|---|---|
| $c_{eq}$ | 2.96 × 10⁻⁶ | 3.017 × 10⁻⁶ | 3.076 × 10⁻⁶ | 3.137 × 10⁻⁶ | 3.2 × 10⁻⁶ | 3.27 × 10⁻⁶ |
| $c_{vis}$ | 2.35 × 10⁻⁴ | 2.96 × 10⁻⁴ | 4.24 × 10⁻⁴ | 7.07 × 10⁻⁴ | 1.57 × 10⁻³ | 6.6 × 10⁻³ |
| $c_{rad}$ | 1.95 × 10⁻² | 1.88 × 10⁻² | 1.79 × 10⁻² | 1.69 × 10⁻² | 1.61 × 10⁻² | 1.57 × 10⁻² |

It is shown in Table III that within the range of glycerol concentration measurement 0-100%, $c_{eq}$ and $c_{vis}$ are much smaller than $c_{rad}$, so the influence of $c_{eq}$ and $c_{vis}$ on (3) can be ignored. Therefore, in our experimental environment, for 0-100% glycerol solutions, the effect of viscosity on amplitude can be ignored.
According to (4), (7), (8), and (13), (3) can be rewritten as,

$$A \approx \frac{F_0}{\sqrt{\left[(\omega_{eq}^2-\omega^2)m_{eq}-0.65\pi\omega^2\rho_f a^3\right]^2+\omega^2\left(\frac{4\pi a^2 c\rho_f(ka)^2}{(1+ka)^2}\right)^2}} \quad (14)$$

Fig. 1(b) shows the amplitude-frequency characteristic curve obtained from the simulation of the sensor immersed in a glycerol solution. From this curve, it can be observed that as the density of the liquid increases within a certain range, the resonant frequency of the PMUT decreases ($\omega_{01}>\omega_{02}>...\omega_{0n}$) [31]. When the resonant frequency of the PMUT corresponding to the unknown liquid to be measured is between $\omega_{01}$ and $\omega_{0n}$, then as long as a single excitation frequency greater than or equal to $\omega_{01}$ is selected, the output voltage of the PMUT corresponding to all liquids within the density range can be uniquely corresponded to the liquid density. For example, in Fig. 1(b), $\omega_m$, which is greater than or equal to $\omega_{01}$, is selected as the single excitation frequency to obtain the voltage value $A\omega_{0i}$ ($i=0\sim n$), which decreases with the increase of liquid density ($A\omega_{01}>A\omega_{02}>...A\omega_{0n}$). Therefore, in a fixed-frequency measurement scheme, a one-to-one relationship exists between the liquid density and the amplitude of the PMUT.

During the propagation of sound waves, the decrease in amplitude is due to the acoustic path (spherical acoustic wave) and the acoustic medium losses due to the viscosity[31]. For such a fixed-distance ultrasonic wave propagation, the viscous attenuation coefficient is $\alpha \approx \frac{2\cdot\pi^2\cdot f_{liquid}^2\cdot\eta}{\rho_f\cdot c_{liquid}^3}$, where $f_{liquid}$ is the frequency of the acoustic wave in the liquid, it is a constant under a fixed frequency measurement; $\eta$ is the viscosity of the liquid; $\rho_f$ is the density of the liquid; and $c_{liquid}$ is the velocity of sound in the liquid; $\rho_f$ is the density of the liquid [31]. Therefore, the signal received by the PMUT receiver decays exponentially due to viscous loss ($e^{-\alpha\cdot AP}$, $AP$ is the propagation distance of ultrasound). As shown in Table IV, the viscous attenuation coefficient is very small, so $e^{-\alpha\cdot AP}$ is close to 1 ($0.99999995 < e^{-\alpha\cdot AP} < 0.999989$). This indicates that in our experiment, the attenuation caused by viscosity during the propagation of sound waves can be ignored.

Based on the theoretical derivation, it can be known that as the density of the liquid increases, the amplitude of the sound wave at the transmitting end of the PMUT diaphragm decreases. At the same time, the amplitude of the sound wave received by the receiving end of the PMUT decreases, and the diaphragm amplitude decreases. Correspondingly, the amplitude of the output voltage at the receiving end decreases.

TABLE IV
SOUND WAVE ATTENUATION COEFFICIENT IN GLYCEROL SOLUTIONS OF DIFFERENT CONCENTRATIONS

| Glycerol solution concentration | 0 | 20% | 40% | 60% | 80% | 100% |
|---|---|---|---|---|---|---|
| attenuation coefficient $\alpha$ | $5.04 \times 10^{-6}$ | $6.25 \times 10^{-6}$ | $9.81 \times 10^{-6}$ | $2.06 \times 10^{-5}$ | $7.73 \times 10^{-5}$ | $1.07 \times 10^{-3}$ |

Fig.1(c) shows the amplitude-frequency characteristic curve of the receiver when the resonance frequencies of the transmitter PMUT and receiver PMUT do not match during the simulation of this experiment. When two PMUTs exhibit a resonance frequency mismatch (manifested as a double peak on the amplitude frequency response curve), the measurement frequency should be greater than or equal to $\omega_m$ shown in Fig. 1(c). In this case, as the density increases, the amplitude of the transmitter PMUT and the output voltage of the receiver PMUT still follow the aforementioned conclusion. Therefore, this fixed-frequency measurement method not only compensates for errors caused by transmitter-receiver PMUT mismatch but also enables density measurement that is decoupled from the effects of viscosity.

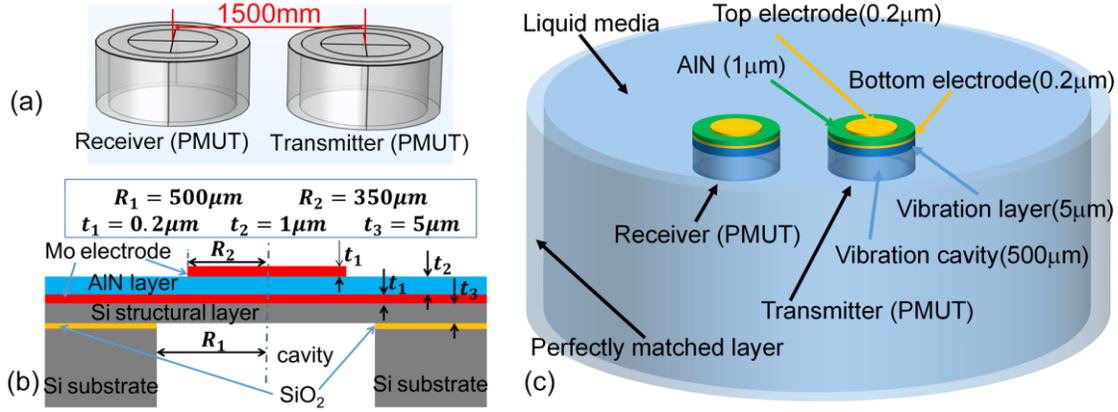

Fig. 2. (a)FEM simulation models of transmitter PMUT and receiver PMUT. (b)Sectional view of the structure of PMUT. (c)Parameters of the PMUT models.

## 3. DEVICE DESIGN AND MODELING

The feasibility of the working principle of the PMUT transmission and reception system was verified using the finite element method (FEM) simulation method. We used a three-dimensional model for modeling. Fig. 2 shows the modeling structure.

As shown in Fig. 2(a), we designed two three-dimensional models of PMUT units, with a distance of 1.5 mm between them, placed in the middle of the bottom of the liquid medium. The PMUT diaphragm consists of the following layers from top to bottom: 0.2 μm top electrode layer (Molybdenum); 1 μm AlN layer; 0.2 μm bottom electrode layer (Molybdenum); 5 μm silicon layer. The top electrode diameter is 700 μm, vibration cavity diameter is 1000 μm. The structure of the PMUT unit is shown in Fig. 2(b). The entire liquid medium is cylindrical, with a circular cross-section at the top serving as the contact surface between the medium and air. Ultrasonic waves reflect through this surface and propagate from the transmitter PMUT to the receiver PMUT. There is a 0.5 mm-thick perfectly matched layer on the sides of the cylindrical liquid medium. Next, we set up the physical fields for the 3D model to facilitate subsequent simulations. The PMUT unit structure is set as a "solid mechanics" physical field; the AlN layer is set as an "electrostatic" and "piezoelectric effect" physical field; the liquid medium is set as a "pressure acoustics" physical field, with the upper cross-section of the medium being the contact surface between the medium and air, which is the sound wave reflection surface. The surrounding area of the medium is a perfectly matched layer for absorbing sound waves, and the other interfaces are set as hard acoustic field boundaries, as shown in Fig. 2(c). Applying an alternating signal with a voltage of 3 V between the electrodes of the PMUT unit at the transmitter, the current density of the liquid medium can be calculated by measuring the voltage amplitude of the output signal between the electrodes of the PMUT unit at the receiver.

We use a fixed non-resonant frequency as the excitation signal frequency at the transmitter to simulate

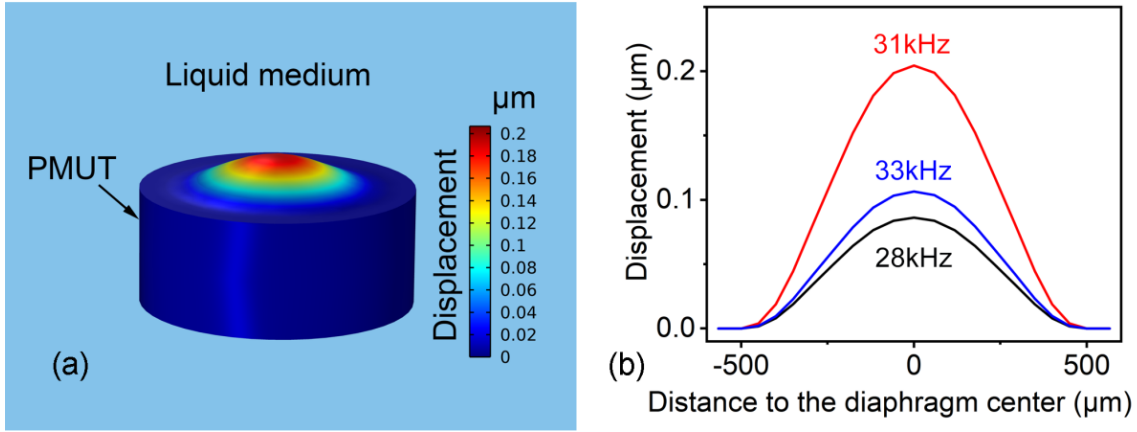

Fig. 3. (a) FEM simulation results of the vibration modes of PMUT at a non-resonant frequency. (b) FEM simulated displacements of the vibration modes of the PMUT at different frequencies.

the vibration state of the PMUT diaphragm at non resonant frequency.

### A. FEM Simulation of the Vibration Modes of PMUT at Non-Resonant Frequency

Fig. 3(a) shows the simulation results of the displacement distribution of the PMUT diaphragm in the medium. It can be seen that the maximum displacement occurs at the center of PMUT, while the displacement of the diaphragm decreases monotonically from the center to the periphery, which can be determined as a typical first-order resonant mode. Fig. 3(b) shows the displacement amplitude of points on the diameter of the diaphragm at the receiver PMUT. The three curves represent the displacement amplitude curves under the excitation of alternating signals at three different frequencies. The red line represents the displacement amplitude curve when the excitation signal is set to the resonant frequency of the device, while the black and red lines represent the displacement amplitude curves when the excitation signal is set to the non-resonant frequency of the device. For all lines, the displacement amplitude increases closer to the center of the diaphragm. From the figure, it can be seen that the vibration of the diaphragm at the receiver PMUT is a first-order vibration mode. This proves that when measuring with non-resonant frequencies, the receiver PMUT will also output an alternating signal.

The simulation results indicate that the combination system of transmitter PMUT and receiver PMUT arranged side by side is theoretically feasible for characterizing liquid density.

### B. Simulation of PMUTs with Different AlN Layer Thicknesses

We first choose a cavity size of 1000 μm for PMUT. Due to the influence of AlN layer thickness on the stiffness and surface density of PMUT thin films, it influences the amplitude and resonant frequency of the transmitter PMUT and the output voltage amplitude of the receiver PMUT. In finite element simulation, we set the medium as pure water with a density of 1 g/cm$^3$. We changed the thickness of the AlN layer and measured the voltage amplitude of the output signal between the electrodes at the receiver PMUT when the excitation signal frequency was the resonant frequency, as shown in Fig. 4(a). The output signal is strongest

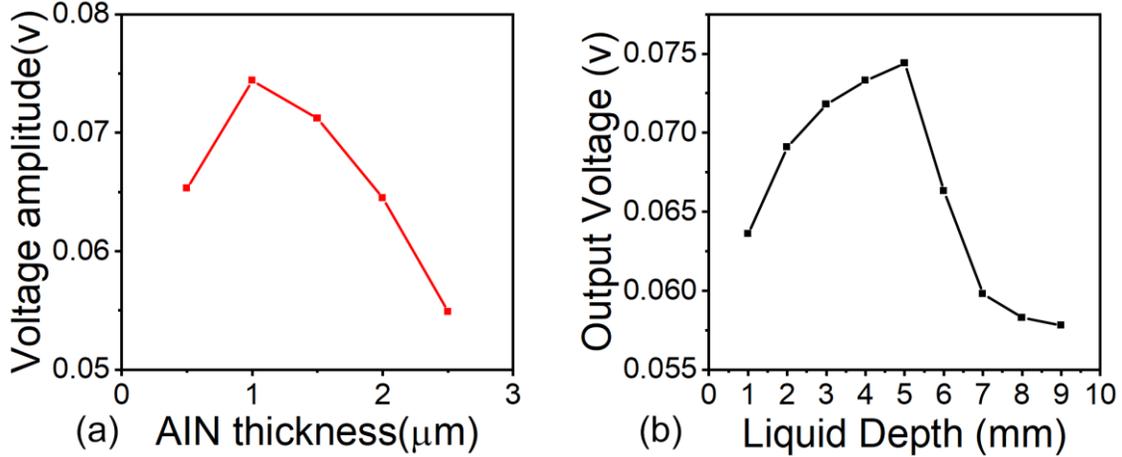

Fig. 4. (a)Simulation results of PMUT with different AlN layer thicknesses. (b) The simulated output voltage amplitude of receiver PMUT in pure water at different depths.

when the thickness of the AlN layer is 1 μm, so we set the thickness of the AlN layer to 1 μm.

C.  Simulation of PMUTs Having Different Cavity Diameters

We set the AlN thickness of two PMUTs to 1 μm, and the cavity diameter to 700, 1000, and 1300 μm, respectively. The medium densities are set to 1, 1.0261, 1.0522, 1.0783, 1.1044, 1.1305, and 1.1566g/cm$^3$, respectively, corresponding to densities of 0-60% glycerol aqueous solution concentration. The resonance frequency range of glycerol aqueous solution with a simulated measurement concentration of 0-60% is shown in the Table V. If we use the maximum frequency in this range as the measurement signal frequency (i.e. the resonant frequency of this PMUT in pure water), and the density of the medium is set to the aforementioned value, then the output voltage range of the receiver PMUT is shown in the fourth row of the Table V. From the data in the table, it can be seen that when the cavity diameter is 1000 μm, using a single measurement signal frequency instead of sweeping frequency measurement can collect voltage signals of appropriate size. When the cavity diameter is 1300 μm, the voltage amplitude of the output signal is high, and the resonance frequency bandwidth is narrow. A small number of fixed measurement frequencies can cover the entire frequency bandwidth, yet the disadvantage is that the Q value of the device is lower at this time. When the cavity diameter is 700 μm, the voltage amplitude of the output signal are not high, and the resonance frequency range is large. More fixed measurement frequencies are required to cover the entire frequency bandwidth, which increases the complexity of the measurement circuit. Therefore, a compromise of choosing a cavity diameter of 1000 μm is preferable.

TABLE V
SIMULATION RESULT OF RESONANT FREQUENCY, Q VALUE, AND OUTPUT VOLTAGE OF PMUTS HAVING DIFFERENT SIZES

| Cavity Diameter (μm) | 700 | 1000 | 1300 |
|---|---|---|---|
| Resonant frequency bandwidth (kHz) | 5.3 | 3.2 | 1.5 |
| Q value | 83.8 | 81.75 | 74.7 |
| Voltage amplitude of the receiver PMUT (V) | 0.03319 | 0.07444 | 0.0945 |

TABLE VI
OUTPUT VOLTAGE VALUES OF PMUTS AT DIFFERENT DISTANCES

| Distance (μm) | 1200 | 1500 | 1800 | 2000 | 2500 | 3000 |
|---|---|---|---|---|---|---|
| Voltage amplitude (V) | 0.0747 | 0.0744 | 0.0717 | 0.0701 | 0.0659 | 0.0444 |

D. Simulation of PMUTs at Different Distances

To investigate the influence of the distance between PMUTs on device performance, we set the distances between the transmitter and receiver PMUTs to 1200, 1500, 1800, 2000, 2500, and 3000 μm, with the output voltages shown in Table VI. It is found that as the distance becomes closer, the amplitude of the output voltage at the receiver PMUT increases. Therefore, while ensuring no transverse wave interference, we determined the distance between the two PMUTs to be 1500 μm.

E. Simulation of PMUTs in Pure Water at Different Depths

To investigate the impact of liquid depth on measurement results, we set the vibration cavity diameter of the PMUT, AlN thickness, and PMUT distance to 1000 μm, 1 μm, and 1500 μm, respectively. The liquid depths are set to 1, 2, 3, 4, 5, 6, 7, 8, and 9 mm. The liquid is pure water. The simulated output voltage amplitude at the receiving end is shown in Fig. 4(b). According to the simulation data, the output signal is strongest when the liquid depth is 5 mm. Therefore, the liquid depth is set to 5 mm.

F. Simulation of Manufacturing-Induced Dimensional Errors on Device Performance

In order to investigate the impact of PMUT size parameter errors caused by manufacturing processes on device performance, we simulated a PMUT device with a cavity diameter of 1000 μm. We changed the cavity radius of the transmitter PMUT by reducing it by 6, 10, 14, 18, and 20 μm, respectively. When the excitation signal frequency in pure water was the resonant frequency of pure water, the output voltages at the receiver PMUT were shown in Fig. 5(a) as 0.0728 V, 0.0692 V, 0.0654 V, 0.0622 V, and 0.0585 V, respectively. When the release cavity radius error of two PMUTs is 20 μm, the output voltage at the receiver PMUT decreases by 21.37% and the Q value decreases by 34.8%. This also proves the importance of manufacturing processes in improving device performance.

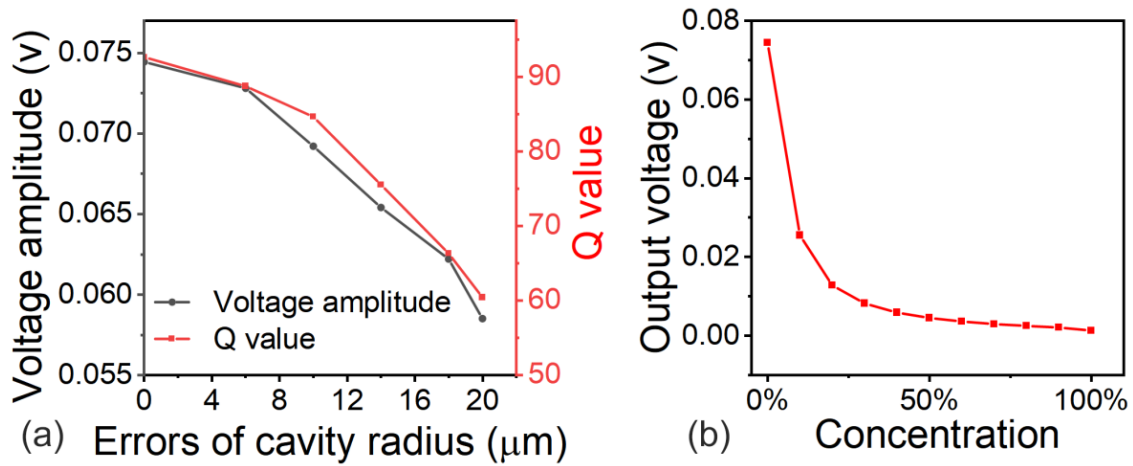

Fig. 5. (a)The impact of size parameter errors caused by manufacturing processes on device performance. (b)Output voltage value of the receiver PMUT in medium with different densities at a fixed frequency of 31.9 kHz.

### G. Simulation of the PMUTs in Medium with Different Densities at A Fixed Frequency

In the simulation, the thickness of AlN layer was set as 1 μm, the cavity diameter was set as 1000 μm, the silicon layer thickness was set as 5 μm, the distance between two PMUT devices was set as 1500 μm and the liquid depth was set as 5 mm. The parameter of the medium was changed from 1 to 1.261 g/cm$^3$ incrementally, which denoted the concentration of the medium changing from 0% to 100%. Fig. 5(b) shows the output voltage of the receiver PMUT at a fixed frequency of 31.9 kHz for media of different densities. With the medium density increasing, the voltage value decreases approximately exponentially.

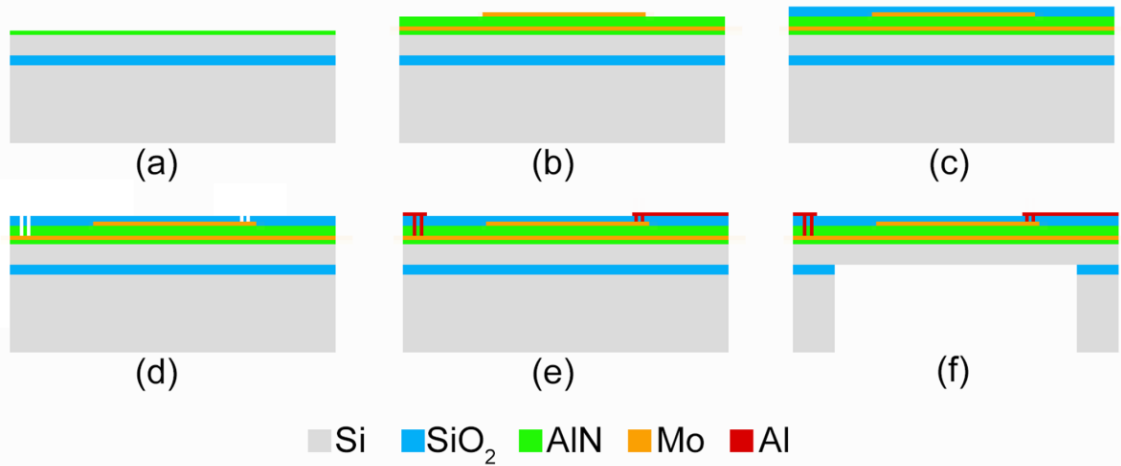

Fig. 6. Fabrication process flow of the PMUT. (a) AlN seed layer depositing. (b) Top electrode paterning. (c) Oxide layer depositing. (d) Top and electrode opening. (e) Al pad patterning. (f) Bottom releasing.

## 4. FABRICATION PROCESS

The fabrication process of the PMUT is shown in Fig. 6. The PMUT is manufactured using a five-mask process on MEMS specific SOI wafer. Firstly, a 0.02 μm AlN seed layer, a 0.2 μm Mo lower electrode layer, a 1 μm AlN piezoelectric layer, and a 0.2 μm Mo upper electrode layer are sputtered onto a MEMS-specific SOI wafer, as shown in Fig. 6(a) and (b). The top circular Mo layer is etched by plasma. Then, using plasma-enhanced chemical vapor deposition (PECVD), a 0.2 μm low-stress silicon dioxide was deposited, as shown in Fig. 6(c) and (d). In Fig. 6(e), when drawing the through holes of the top and bottom electrodes, the silicon dioxide layer was etched using reactive ion etching (RIE) with CHF3, and the AlN layer was etched using

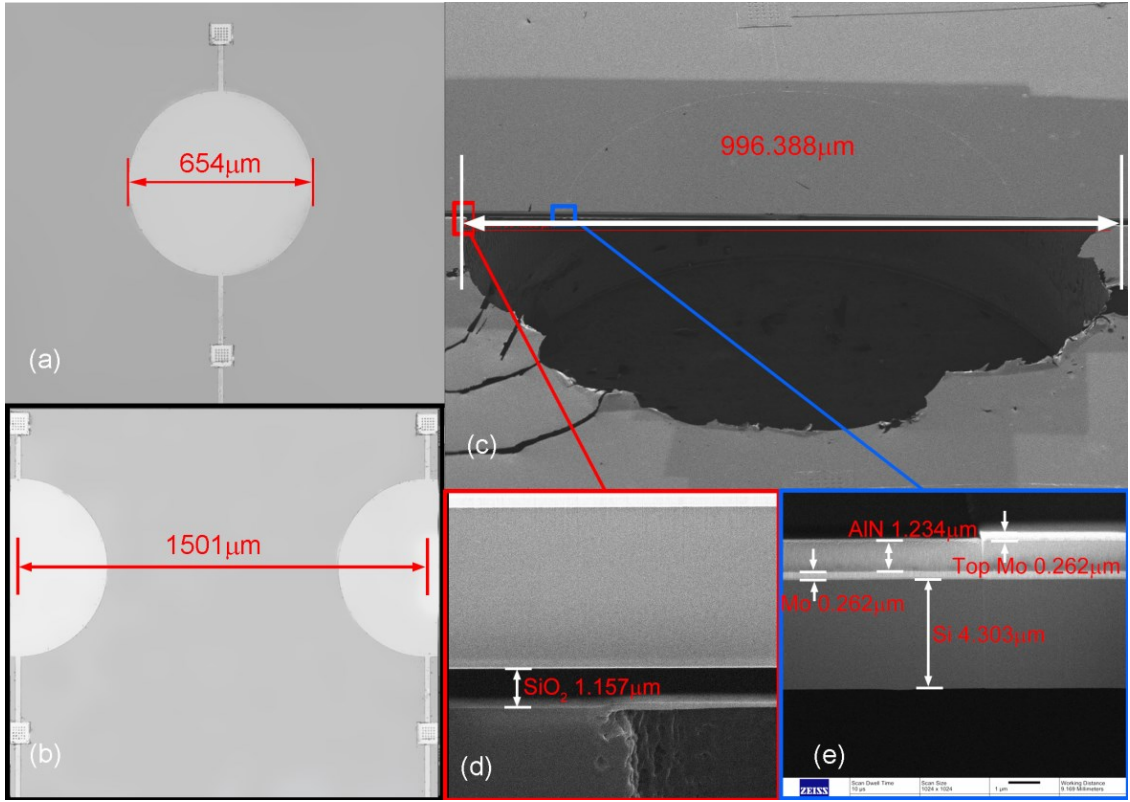

Fig. 7. (a) and (b) Microscope image of the PMUT. (c) Tilted HIM image of the cut membrane and exposed cavity. (d) Zoomed-in view HIM image of the boundary of back-etched cavity. (e) Zoomed-in view HIM image of the cross-sectional structure of the released membrane.

RIE combined with chlorine gas and boron trichloride gas. Then, deposit a 0.7 μm Al layer and etch it into bonding pads. Finally, deep reactive ion etching (DRIE) and buffered oxide etching (BOE) were performed on the backside of the MEMS specific SOI wafer to remove the Si and $SiO_2$ layers, releasing the PMUT film, as shown in Fig. 6(f). Microscope images of the PMUTs are shown in Fig. 7. Fig. 7(a) and (b) present microscopic images of the PMUT, showing a top electrode diameter of 654 μm and a center-to-center distance of 1501 μm between two adjacent PMUTs. Fig. 7(c) displays a cross-sectional image captured by tilted helium ion microscopy (HIM). The membrane layer was released and the backside etched cavity was exposed through gallium-based focused ion beam (FIB) milling, with the cavity diameter measured as 996.388 μm. Enlarged views of the cross-sectional structure of the PMUT membrane in Fig. 7(d) and (e) clearly reveal the experimentally determined thicknesses of each layer.

## 5. RESULTS AND DISCUSSION

### A. Experiment Setup

The setup of the PMUT liquid density sensor characterization is shown in Fig. 8(a). The PCB board with PMUTs is glued to the bottom of the culture dish, and the electrodes of the transmitter PMUT and receiver PMUT are connected to the PCB board through electrical wires. The PCB board is connected to the measurement setup via a coaxial cable. An appropriate amount of glycerol solution is added to the culture dish as a medium.

We use a phase-locked amplifier (MFLI 500, Zurich Instrument, Switzerland) to measure devices in liquid media. As shown in Fig. 8(a), the signal output port of this phase-locked amplifier is connected to the two electrodes of the transmitter PMUT and outputs an alternating excitation signal. The signal input port of this phase-locked amplifier is connected to the output end of the SR560 signal amplifier, and the input end of the SR560 signal amplifier is connected to the two electrodes of the receiver PMUT. In this way, the alternating electrical signal generated by the receiver PMUT will pass through SR560 filtering, amplification(20 times), and be input into the phase-locked amplifier, where the voltage amplitude of this alternating signal is obtained.

Theoretically, a single frequency is sufficient to distinguish the output voltage of sensors in different densities. As a result, we measure the vibration displacement value of the PMUT receiver in the glycerol solution with a liquid concentration range of 0–100% when the input signal is a single frequency using digital holography microscopy (DHM-R2100, Lyncée Tec SA, Switzerland) as shown in Fig. 8(b).

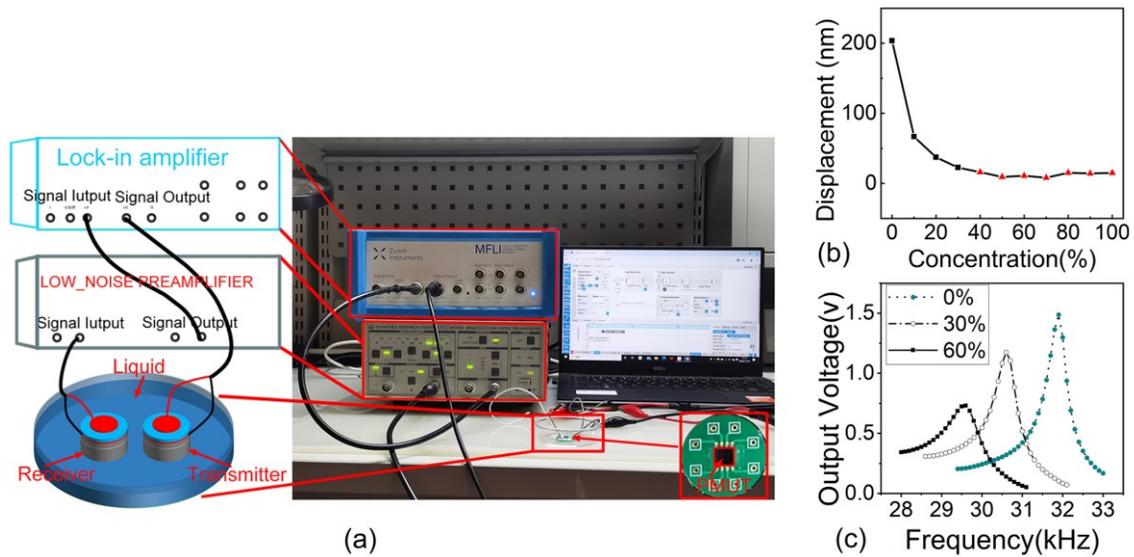

Fig. 8. (a) Experimental setup of the PMUTs liquid density sensor. (b) Vibration displacement value of the PMUT output end in the liquid with a concentration range of 0% to 100% under the excitation signal frequency of the resonance frequency at 0% concentration by DHM lens. (c) The signal (peak-to-peak amplitude) received at the receiver in the three fluids of concentrations 0, 30% and 60%.

However, the experimental results show that when the excitation signal of the PMUT transmitter is set to the resonance frequency corresponding to 0% glycerol solution, the output voltage value of the PMUT receiver will gradually decrease when the liquid concentration changes from 0% to 30%. When the concentration of the glycerol solution rises from 30% to 100%, the output voltage value is too small and easily affected by noise. As a result, we cannot accurately measure the output voltage. Therefore, if the excitation signal is set to a single frequency, the DHM, as a high-resolution device, can only detect the glycerol solution with a concentration range of 0-30%. This proves that the actual experimental range of single-frequency measurement of liquid concentration is narrower than the theoretical value. In order to expand the measurement range, we adopted a three-frequency measurement method to measure the liquid density.

## B. Experiments with 0-100% Concentration Glycerol Solution

According to the working principle, the fixed measurement frequency must be greater than the resonance frequency of all the tested liquids. For example, if the concentration range of the glycerol solution is 0-30%, the resonance frequency decreases with the increase of concentration when the PMUT vibrates in the liquid within its concentration range. Therefore, the excitation frequency we selected should be greater than or equal to the resonance frequency of the PMUT in a 0% concentration glycerol solution. We begin by categorizing the test liquid—a glycerol solution ranging from 0% to 100% (corresponding to a density range of 1–1.261 g/cm³)—into three concentration intervals: 0–30%, 30–60%, and 60–100%. A fixed measurement frequency was determined for each range. To accomplish this, frequency scanning was performed on the PMUTs to identify their resonance frequency in three reference solutions with glycerol concentrations of 0%, 30%, and 60%. The peak-to-peak amplitude of the signal received by the receiver is shown in Fig. 8(c). Based on these results, the three frequencies—31.9 kHz, 30.6 kHz, and 29.6 kHz, which are resonant frequencies of PMUTs in pure water, 30% glycerol solution, and 60% glycerol solution—were selected as the designated measurement frequencies for each respective concentration range.

For glycerol solutions ranging from 0% to 30% (corresponding to a density range of 1–1.0783 g/cm³), the resonant frequency of PMUTs in pure water (31.9 kHz) was selected as the measurement frequency to determine the density of the glycerol solutions. This ensures that the fixed measurement frequency remains higher than the resonant frequency of all liquids under test. Sixteen liquid samples, with densities of 1 (0%), 1.0052 (2%), 1.0104 (4%), 1.0157 (6%), 1.0209 (8%), 1.0261 (10%), 1.0313 (12%), 1.0365 (14%), 1.0418 (16%), 1.047 (18%), 1.0522 (20%), 1.0574 (22%), 1.0626 (24%), 1.0679 (26%), 1.0731 (28%), and 1.0783 g/cm³ (30%), were prepared with concentrations evenly spaced between 0% and 30%. The output voltage of the device in each liquid was measured at a frequency of 31.9 kHz, and the relationship between liquid density and output voltage was plotted. Since the output voltage amplitude decays exponentially with increasing liquid density, an exponential decay function was used to fit the data, as shown in Fig. 9(a). The resulted fitting equation for calculating the density of glycerol solutions within this density range is as follows:

$$\rho = 0.95107 + 0.08847 \times e^{-\frac{x}{0.19579}} + 0.07362 \times e^{-\frac{x}{3.62561}} \quad \left(1\frac{g}{cm^3} \le \rho < 1.0783\frac{g}{cm^3}\right) \quad (15)$$

where $x$ is the measured output voltage value, measured in volts.

For glycerol solutions ranging from 30% to 60% (corresponding to a density range of 1.0783–1.1566 g/cm³), the resonant frequency of PMUTs in a 30% glycerol solution (30.6 kHz) was selected as the measurement frequency to determine the density within this range. This ensures that the fixed measurement frequency remains higher than the resonant frequency of all liquids under test. Similarly, sixteen liquid

samples with densities of 1.0783 (30%), 1.0835 (32%), 1.0887 (34%), 1.0940 (36%), 1.0992 (38%), 1.1044 (40%), 1.1096 (42%), 1.1148 (44%), 1.1201 (46%), 1.1253 (48%), 1.1305 (50%), 1.1357 (52%), 1.1409 (54%), 1.1462 (56%), 1.1514 (58%), and 1.1566 (60%) g/cm³, were prepared with concentrations evenly spaced between 30% and 60%. The output voltage of the device in each liquid was measured at a frequency of 30.6 kHz, and the relationship between liquid density and output voltage was plotted. Since the output voltage amplitude decays exponentially with increasing liquid density, an exponential decay function was used to fit the data, as shown in Fig. 9(b). The resulted fitting equation for calculating the density of glycerol solutions within this range is as follows:

$$\rho = 4614.059 \times e^{-\frac{x}{113355.113}} + 0.08397 \times e^{-\frac{x}{0.15987}} - 4612.9818$$

$$\left(1.0783 \frac{g}{cm^3} \leq \rho < 1.1566 \frac{g}{cm^3}\right) \quad (16)$$

For glycerol solutions ranging from 60% to 100% (corresponding to a density range of 1.1566–1.261 g/cm³), the resonant frequency of PMUTs in a 60% glycerol solution (29.6 kHz) was selected as the measurement frequency. Following the same procedure as described above, twenty-one liquid samples with densities of 1.1566 (60%), 1.1618 (62%), 1.167 (64%), 1.1723 (66%), 1.1775 (68%), 1.1827 (70%), 1.1879 (72%), 1.1931 (74%), 1.1984 (76%), 1.2036 (78%), 1.2088 (80%), 1.214 (82%), 1.2192 (84%), 1.2245 (86%), 1.2297 (88%), 1.2349 (90%), 1.2401 (92%), 1.2453 (94%), 1.2506 (96%), 1.2558 (98%), and 1.261 g/cm³ (100%), were prepared with concentrations evenly spaced between 60% and 100%. The output voltage of the device in each liquid was measured at a frequency of 29.6 kHz, and the relationship between liquid density and output voltage was plotted. Since the output voltage amplitude decays exponentially with increasing liquid density, an exponential decay function was used to fit the data, as shown in Fig. 9(c). The resulted fitting equation for calculating the density of glycerol solutions within this range is as follows:

$$\rho = 1.17245 + 0.11159 \times e^{-\frac{x}{0.18459}} - 0.0001616 \times e^{-\frac{x}{0.15525}}$$
$$\left(1.1566 \frac{g}{cm^3} \leq \rho \leq 1.261 \frac{g}{cm^3}\right) \quad (17)$$

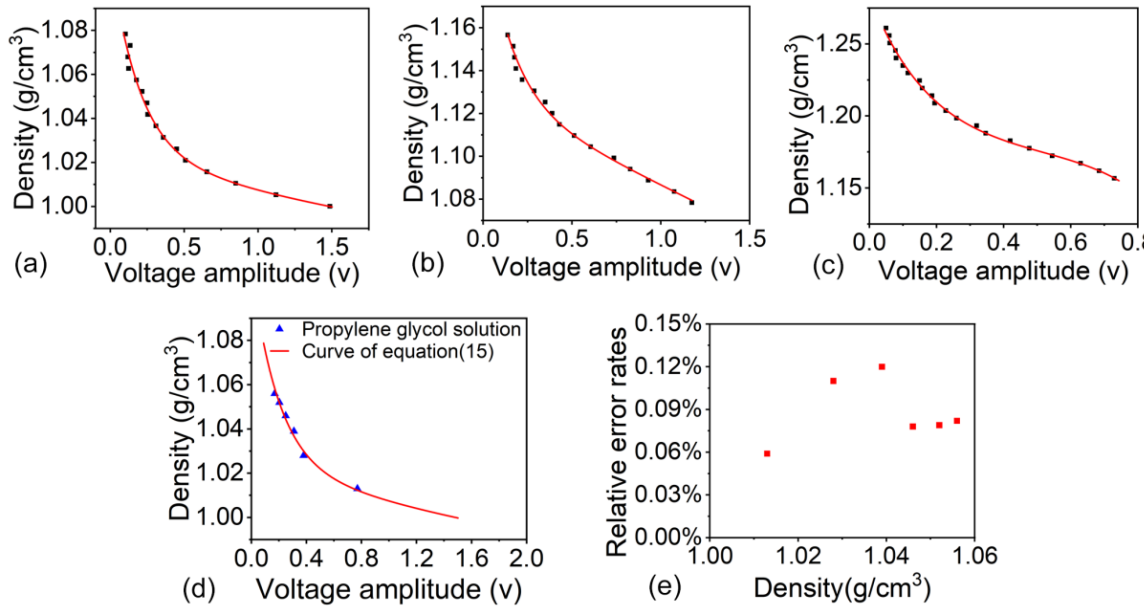

Fig. 9. (a)The output voltage values from the concentration range of 0-30% are fitted with an exponential decay function. (b)The output voltage values from the concentration range of 30%-60% are fitted with an exponential decay function. (c)The output voltage values from the concentration range of 60%-100% are fitted with an exponential decay function. (d) The output voltages of the six propylene glycol solutions closely align with the sensor's calibration curve obtained from glycerol solution in section 5B, which was established using glycerol solutions. (e) Measured density error rate for propylene glycol solutions.

Thus, the calibration of the liquid density sensor is completed.

After the calibration of the density sensor is completed, a three-segment curve can be used to measure the glycerol solution with unknown density. The specific strategy is as follows: first, use the measurement frequency of 31.9 kHz to measure the glycerol solution with an unknown concentration. If the density calculated by (15) meets 1 g/cm³≤$\rho$<1.0783 g/cm³ (corresponding to the concentration of glycerol solution is 0%-30%), it indicates that the measurement result has been obtained and the measurement ends. Otherwise, continue to use the measurement frequency of 30.6 kHz and calculate the liquid density by (16). If the result satisfies 1.0783 g/cm³≤$\rho$<1.1566 g/cm³ (corresponding to the concentration of glycerol solution is 30%-60%), it indicates that the measurement result has been obtained and the measurement ends. Otherwise, continue to use the measurement frequency of 29.6 kHz and calculate the liquid density by (17); at this time, the measurement result has been obtained. After running the above process, the measurement ends.

## C. Verification of the insensitivity to liquid viscosity variations

To verify that the viscosity has negligible effects on liquid density measurement with the method presented in this paper, the liquid density sensor calibrated with glycerin solution is used to measure the density of propylene glycol aqueous solutions. Six propylene glycol solutions are prepared with their densities covering the range from 1.013 g/cm³ to 1.056 g/cm³. As shown in Table VII, propylene glycol solutions of the same density present significantly different viscosities compared with glycerin solutions. For example, with the same density of 1.056 g/cm³, the viscosity of propylene glycol solution is 21.04 cP, which is 11.19 times larger than that of glycerin solution. Following the measurement procedure, the excitation frequency is fixed at 31.9 kHz to measure the output voltages of the PMUT receiver in propylene glycol solutions. To reduce the experimental errors as much as possible, the voltage measurement is repeated three times and averaged. As shown in Fig. 9(d), the red line is the sensor calibration curve established with glycerin solutions of 0–30% concentrations, and the six triangle data points are the mean values of the voltages output by the PMUT receiver in six propylene glycol solutions. It can be seen that these six data points coincide substantially with the calibration curve. Substituting the output voltage values into (15), the error rates between the calculated densities and the true densities are shown in Fig. 9(e). The results show that for different types of liquids, as long as their density is within the measurable range of the sensor, accurate measurement can still be achieved even if there exist differences in viscosity. This verifies that for the liquid density sensor operating at a fixed excitation frequency, the effect of viscosity variation on density measurement can be neglected. As a result, the calibration curve of (15) applies not only to glycerin, but also to other types of liquids within the liquid density range.

TABLE VII
COMPARISON OF VISCOSITY BETWEEN PROPYLENE GLYCOL SOLUTION AND GLYCEROL SOLUTION WITH THE SAME DENSITY

| Density (g/cm³) | Viscosity(cp) | |
|---|---|---|
| | Propylene glycol solution | Glycerol solution |
| 1.013 | 1.83 | 1.08 |
| 1.028 | 3.06 | 1.31 |
| 1.039 | 5.25 | 1.56 |
| 1.046 | 10.04 | 1.68 |
| 1.052 | 14.15 | 1.76 |
| 1.056 | 21.04 | 1.88 |

### D. Measuring the Density of an Unknown Liquid

To measure the density of a liquid with unknown density, follow the procedure below, as illustrated in Fig. 10:

1) Set the excitation frequency to 31.9 kHz (this measurement frequency corresponds to the density range of 1–1.0783 g/cm³) and measure the output voltage of the receiver PMUT in the unknown liquid.

2) Input the measured voltage into (15) corresponding to the density range of 1–1.0783 g/cm³. As shown in section V(B), (15) applies not only to glycerin, but also to other types of liquids within the liquid density range. If the calculated density falls within the range of 1–1.0783 g/cm³, this value represents the density of the unknown liquid. Otherwise, proceed to the next step.

3) Set the excitation frequency to 30.6 kHz (this measurement frequency corresponds to the density range of 1.0783–1.1566 g/cm³) and measure the output voltage of the receiver PMUT in the unknown liquid.

4) Input the measured voltage into (16) corresponding to the density range of 1.0783–1.1566 g/cm³. If the calculated density falls within the range of 1.0783–1.1566 g/cm³, this value is the density of the unknown liquid. Otherwise, proceed to the next step.

5) Set the excitation frequency to 29.6 kHz (this measurement frequency corresponds to the density range of 1.1566–1.261 g/cm³) and measure the output voltage of the receiving PMUT in the unknown liquid.

Input the measured voltage into (17) corresponding to the density range of 1.1566–1.261 g/cm³. If the calculated density falls within the range of 1.1566–1.261 g/cm³, this value is the density of the unknown liquid. Otherwise, the density of the unknown liquid is beyond the measurable range (1–1.261 g/cm³).

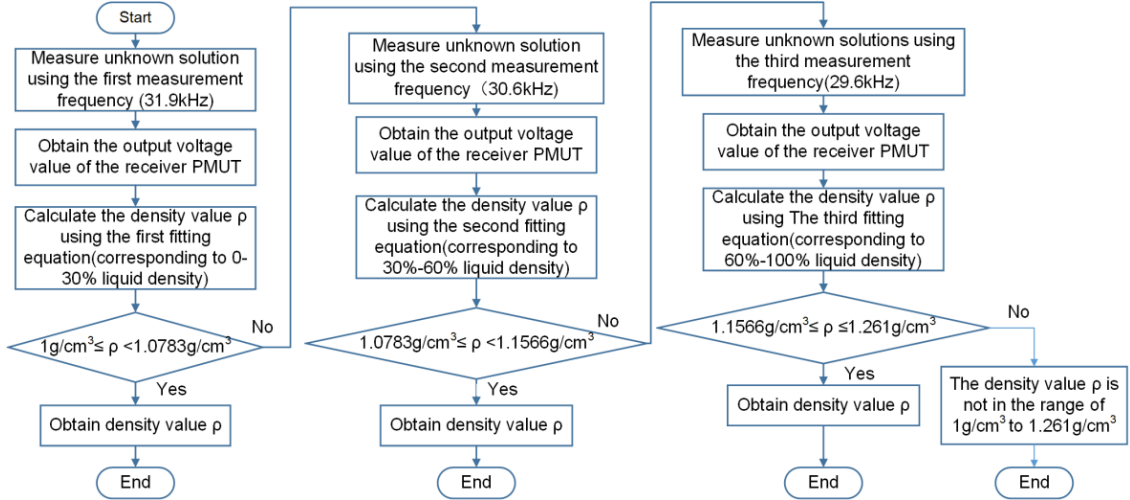

Fig. 10. Flowchart for measuring density using three fixed measure-ment frequencies.

## E. The Influence of Errors Generated in Device Manu-facturing on Density Measurement

Due to inevitable errors between the PMUT transmitter and PMUT receiver during the device manufacturing process, these errors may affect density measurements. We use these 5 devices to measure their output voltage amplitude and Q value in pure water. The measurement results are shown in Fig. 11. The measurement results indicate that the manufacturing process has an impact on device consistency within 5%.

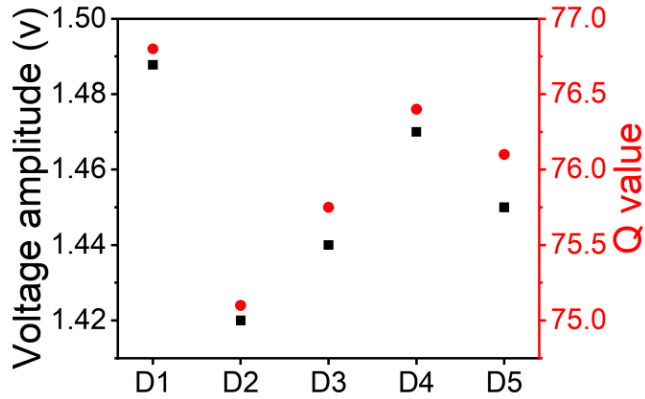

Fig.11. Output voltage amplitude and Q value of 5 devices with the same specifications in pure water.

## F. Comparison With Frequency Sweep Mode

In order to investigate the measurement limit of the method in this paper, the glycerol solution was further diluted to test the variation of the density of the liquid obtained by PMUT under the method with a smaller concentration difference. For this purpose, six concentrations: 0%, 0.1%, 0.25%, 0.5%, 1%, and 2% were selected to perform the tests. At each concentration, five consecutive measurements were performed.

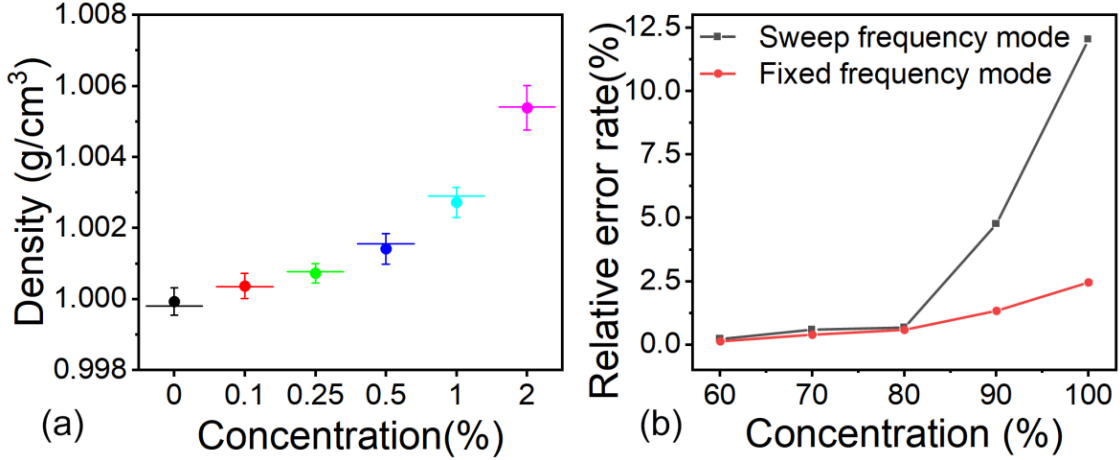

Fig. 12. (a) Density fluctuation measured by the PMUT density sensor in 0%-2% glycerol solution. (b) Comparison of the error rate of density measurement using fixed frequency mode and sweep frequency mode for this PMUT density sensor in high concentration glycerol solution.

Fig. 12(a) shows the density values measured by the method in this paper as a function of concentration from 0 to 2% for the aqueous solution of glycerol. By averaging the multiple measurements, the average density shows a difference, and it is possible to distinguish the concentration changes. In the lowest concentration change, the average density value between pure water and 0.1% glycerol solution increases from 1 to 1.00026 g/cm³. It can be assumed that the resolution of measuring the concentration of glycerol by the method is 0.1%, where the detection limit of density can be calculated as

$$\text{Resolution} = \rho_{0.1\%glycerol} - \rho_{purewater} = 2.6 \times 10^{-4} \frac{\text{g}}{\text{cm}^3} \tag{18}$$

where $\rho_{0.1\%glycerol}$ is the density of 0.1% glycerin, and $\rho_{purewater}$ is the density of pure water. The measurement results show that this method has high resolution to measure the liquid density as small as 2.6×10⁻⁴ g/cm³.

Frequency sweep mode first finds the resonant frequency corresponding to each density, and then calculates the density based on the linear relationship between the resonant frequency and density. However, in high viscosity liquids, this linear relationship does not exist due to the influence of viscosity, so this method cannot measure the density of high viscosity liquids. We measured high viscosity glycerol solution using this method and the fixed frequency measurement method, and the error rates of the two methods are shown in Fig.12(b), indicating a significant difference between the two at high concentrations.

## G. Hysteresis of the PMUT Density Sensor

To test the hysteresis characteristic of the PMUT-based liquid density sensor, the liquid in the PMUT tank is first filled with pure water. Subsequently, pure glycerol solution is gradually added to the tank to increase the concentration of glycerol until the tank is full of glycerol. After that, pure water is gradually added to the tank to decrease the concentration of glycerol until the tank is full of pure water. The received signals under different concentrations of glycerol are measured and converted into density values, as shown in Fig. 13(a). The experimental results show that the measured density values are basically unchanged during the process of increasing and decreasing the concentration of the liquid, proving that the device has no significant hysteresis effect.

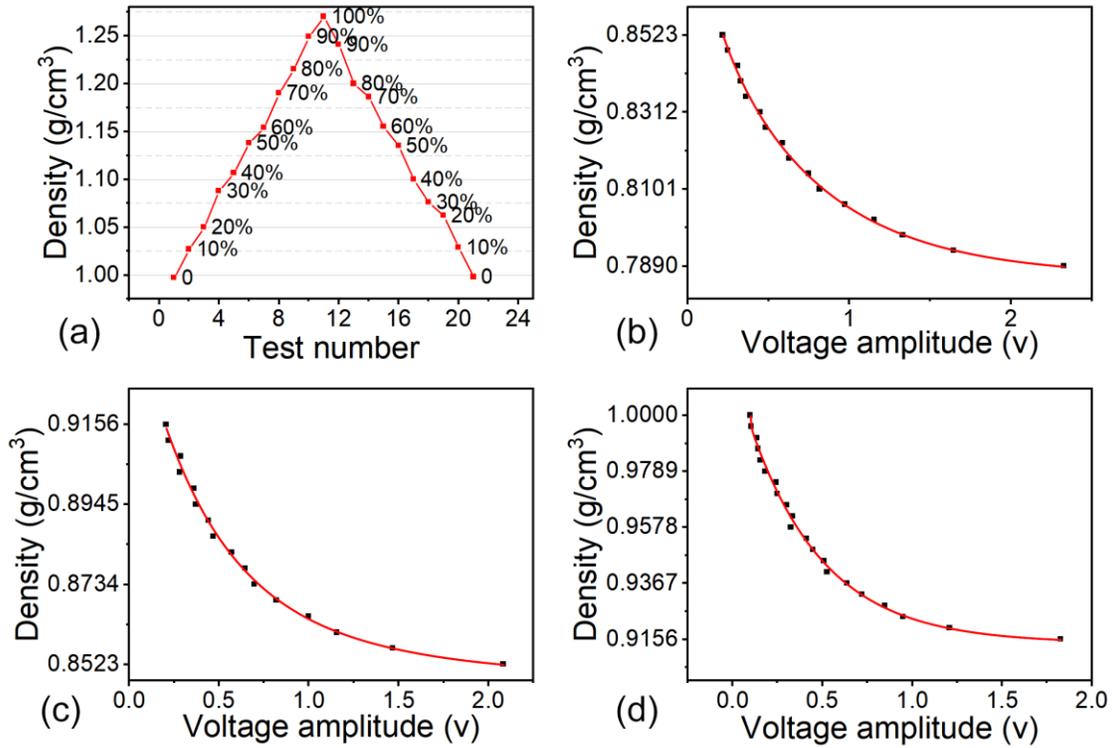

Fig.13. (a) Hysteresis test result of the PMUT density sensor. (b)The output voltage values of alcohol solutions within the concentration range of 100%-70% are fitted using an exponential decay function. (c)The output voltage values of alcohol solutions within the concentration range of 70%-40% are fitted using an exponential decay function. (d)The output voltage values of alcohol solutions within the concentration range of 40%-0 are fitted using an exponential decay function.

### H. Measure the Density of An Alcohol Solution

In order to show that our measurement method can measure liquid density besides glycerin solutions and propylene glycol solutions, we measure alcohol solution density with the three-frequency measurement method mentioned above. The three measurement frequencies are 35.4 kHz, 34.35 kHz, and 33.3 kHz, which are the resonance frequencies of 100%, 70%, and 40% alcohol solution concentrations respectively. The measurement strategy still starts measuring with a larger measurement frequency, the first frequency is used to measure the density range of 0.789-0.8523 g/cm³, the second frequency is used to measure the density range of 0.8523-0.9156 g/cm³, the third frequency is used to measure the density range of 0.9156-1g/cm³. The fitting curve is shown in Fig. 13(b)(c)(d). Using the three fitting equations obtained from these fitting curves, the density of the alcohol solution can be calculated according to the process shown in Fig. 10.

## 7. Conclusion

In this work, we introduce an AlN-based PMUT liquid density sensor only sensitive to liquid density and not sensitive to liquid viscosity variations. Two identical PMUTs placed side by side in the liquid are selected to constitute the transmitter and receiver of the liquid density sensor. A single-frequency excitation voltage is applied to drive the transmitter PMUT to generate ultrasonic waves, which are reflected by the liquid surface and subsequently received by the receiver PMUT to produce an electrical signal. Theoretical calculations demonstrate that when the excitation frequency remains constant, the amplitude of the received electrical signal exhibits a specific relationship with liquid density, while the viscosity-induced signal amplitude variation becomes negligible. Therefore, after calibrating the PMUT liquid density sensor by fitting the relationship between received signal amplitude and density across 0-100% glycerol solutions, the device can measure densities within this range regardless of liquid type and viscosity. In the liquid density range from 1.013 g/cm$^3$ to 1.056 g/cm$^3$, the output voltages of six different density propylene glycol solutions are highly consistent with the calibration curve established by glycerol solution in this density range, and substituting the output voltage value into the fitting equation to calculate the density, the error rate between the calculated density and the real density is smaller than 0.125%. The proposed method achieves maximum error rates less than 2.5% in high-viscosity environments (80%-100% glycerol solutions), which is 20% that of other density measurement methods based on resonant frequency. The PMUT fluid density sensor has a Q value of 76.8 in pure water, and the output signal SNR is up to 126 dB. The detection resolution is $2.6\times10^{-4}$ g/cm$^3$. The proposed PMUT fluid density sensor has advantages in cost, measurable density range, and reliability, and is expected to be applied to the detection of physiological indicators, concentration or uniformity measurement in the chemical industry, and the automatic control field.

## Author contributions

H.R. conceived and designed the study. X.F. performed the experiments and analyzed the data. H.R., X.F., L.L. contributed to the discussions of the experimental results. The paper was written by X.F. and H.R. with input and contributions from L.L. The paper was reviewed and edited by H.R. and L.L. The project was supervised by H.R.